# The role of the observer in the Everett interpretation

H. D. Zeh – www.zeh-hd.de (arxiv:1211.0196v6)

**Abstract:** The role attributed to the observer in various interpretations of quantum mechanics as well as in classical statistical mechanics is discussed, with particular attention being paid to the Everett interpretation. In this context, the important difference between physical properties appearing as "quasi-classical" (robust against decoherence) and "macroscopically given" (rather than being part of a thermodynamic ensemble of states) is pointed out.

**1. The observer in traditional quantum mechanics**

When Werner Heisenberg discovered his matrix mechanics, which denies the general existence of definite values for microscopic physical variables, such as position and momentum of an electron, before they are measured, he invented a historically unprecedented role for "human observers". He assumed that properties of microscopic objects are *created* in an irreversible act of observation – for him confirmation of the superiority of an idealistic world view instead of materialism, realism and reductionism.[1] In particular, the concept of time (including its arrow) would become a fundamental extraphysical prerequisite for the formulation of this process as well as of other physical laws. This point of view was soon supported by his friends Wolfgang Pauli and Carl Friedrich von Weizsäcker. It seemed to become even strengthened when Heisenberg's early attempts to understand his uncertainty principle simply as a consequence of unavoidable perturbations of the electron during measurements (for example by means of his "electron microscope") failed as a consistent explanation.

Niels Bohr later subscribed to a slightly different position by assuming that the outcome of a measurement is *objectively* created as a classically describable property in the measurement apparatus. This picture appears more realistic from the modern description in terms of decoherence, but Bohr rejected all attempts to analyze a measurement as a dynamical process, since he also denied any observer-independent microscopic reality in order to avoid certain consistency problems. Such a creation or coming-into-being of physical properties in measurements "outside the laws of nature" (in Pauli's words) would give the trivial statement that "unperformed experiments do not have any results" a non-trivial meaning. This "quantum creationism" of particle and other observed properties would not only mean that "the click in the counter occurs out of the blue … and without being causally preceded by a decay event in the atom",[2] but also require the possibility of genuine superluminal quantum teleportation in order to explain corresponding experiments. Although such a *strategic position for arguing* may sound ridiculous, it is still widely accepted and used by practicing physicists.

When Erwin Schrödinger proposed his wave equation, he initially expected the single particle wave function he had postulated to describe the real electron, and thus to explain the uncertainty between its apparently observed position and momentum variables by means of the Fourier theorem. However, he could not understand the apparently observed quantum jumps between different standing waves that describe the observed energy eigenvalues. For quite some time he tried in various ways to deny the reality of entangled many-particle wave functions – although precisely this entanglement offers an explanation of apparent jumps by means of decoherence (as we know today). When Max Born invented his interpretation of wave functions as probability amplitudes for the occurrence of traditional particle properties, entanglement was usually but insufficiently interpreted as representing no more than a statistical correlation. So Heisenberg later spoke somewhat mystically about the wave function as representing "human knowledge as an intermediary kind of reality" – an idea that seems to have been revived in the recently quite popular (but similarly insufficient) information theoretical approach to quantum theory. This "quantum Bayesianism" is the most recent form of a shut-up-and-calculate mentality.

Meanwhile, John von Neumann, in his book about the Mathematical Foundations of Quantum Mechanics, had studied the possibility of describing the measurement process as a quantum mechanical interaction between object and apparatus. In strong contrast to Bohr's view, he represented states of a macroscopic "pointer" by quantum mechanical wave packets rather than in classical terms. The unitary interaction, when applied to initial microscopic superpositions, would then lead to entangled superpositions of macroscopically distinct pointer positions and different values of the measured particle property, for example. Therefore, von Neumann postulated a stochastic collapse (or "reduction") of the wave function as a new kind of dynamics supplementing the Schrödinger equation.[*] He called it a "first intervention" – presumably since at that time mainly energy eigenstates were regarded as representing physical states, while their dynamics appeared to consist solely of stochastic quantum jumps between them. The time-dependent Schrödinger equation (his "second intervention") was then mainly used to calculate probabilities for such jumps, although it can also describe deterministically moving wave packets. The collapse proposal came close to Born's original version of his probability interpretation, which was meant to postulate probabilities for jumps from initial to final *wave functions*, which were identified with corresponding eigenfunctions of formal observables. Only after Pauli's intervention had it been re-interpreted as describing probabilities for the creation of classical particle properties, as envisioned by Heisenberg.

---

[*] Heisenberg and Bohr never accepted this dynamical interpretation, but instead regarded the collapse into an individual state as a "normal increase of information" about properties that they assumed to be stochastically created "outside the laws of physics" to form an *ensemble* of potential outcomes (rather than a superposition).

However, von Neumann did not assume his measurement process to end with the apparatus, since he also included the human observer as a quantum system. In this way, the collapse became essential for him in order to re-establish a traditional "psycho-physical parallelism", which seems to be ruled out if the observer remained entangled with the apparatus to form one common state only. Although this description is still compatible with an ontic interpretation of the wave function, the dynamical collapse could be applied anywhere along the "indivisible chain of interactions between the observer and the observed" (von Weizsäcker's words). This free choice is also discussed as a variable position of a conceptual "Heisenberg cut" between microscopic and macroscopic description. For Heisenberg, this freedom was fundamental for his interpretation of quantum theory,[3] while Bohr preferred to apply the probability interpretation at some not very precisely specified border line that would separate the microscopic and the macroscopic realms somewhere within the measurement apparatus.

The conscious observer was further discussed as the key element of the quantum measurement process by London and Bauer.[4] Eugene Wigner even suggested explicitly the possibility of an active influence of consciousness on the physical world,[5] but dropped this proposal when he learned about the concept of what was later called decoherence.[6] Occasional attempts to confirm Wigner's original proposal in the form of deviations from Born's rule caused by the observer's mind indeed failed. Quantum indeterminism thus seems to have nothing to do with an apparent "free will" (as had been hoped for by Born himself). While I therefore prefer to understand von Neumann's "psycho" part in his parallelism in the sense of a passive epiphenomenon, Max Jammer compared it with Anaxagoras' dualistic doctrine of Matter and Mind when he quoted him (adding his own suggested modern interpretation in parentheses):[7] "The things that are in a single world are not parted from one another, not cut away with an axe, neither the warm from the cold nor the cold from the warm" (superposi-tions!?), but "when Mind began to set things in motion, separation took place from each thing that was being moved, and all that Mind moved was separated" (reduction!?).

Other interpretations of quantum mechanics, such as Bohm's or collapse theories, are often claimed *not* to require an observer as an essential part. Although the observer does indeed not assume a specific or extra-physical role in these theories (as he does in the Copenhagen interpretation), John Bell pointed out that Bohm's theory is tacitly based on the assumption of an observer being ultimately described solely by the "classical" and thus local variables that are here *just postulated* to exist in addition to the non-local wave function (but to remain otherwise unobservable).[8] Collapse theories, on the other hand, would not only have to be able to explain the occurrence of quasi-classical narrow wave packets for all macroscopic variables, but also definite states of the conscious observer system in the human brain in order to eliminate superpositions of different states of awareness.

## 2. The observer in classical statistical physics

The observer has always played an essential role in the empirical sciences, simply because the latter are based on observations performed by humans by means of chains of physical interactions with the observed. This remark may appear trivial, but its consequences are non-trivial for all physical concepts that depend on "incomplete information", in particular in statistical mechanics.[9] Why do we regard the position and shape of a solid body as "physically given" even when we do not know them, while we describe the molecules of a gas objectively by an ensemble of their different microscopic states, for example? Since in a Laplacean world all variables are equally real, any such distinction must be based on the vaguely defined difference between what *we* can easily observe and what requires some instrumental effort to find out. Although this distinction is often based on some dynamical stability in contrast to rapid and uncontrollable change, it represents exact conservation laws only in some cases. Stability is certainly relevant, but the boundary separating these realms may vary – for example when we decide to take explicitly into account local fluctuations of quantities out of their equilibrium, or, more drastically, during phase transitions, when new order parameters may form.

The dependence of thermodynamical concepts on incomplete knowledge is particularly obvious in Willard Gibbs' approach to statistical thermodynamics, which is defined in terms of $\Gamma$-space distributions that are related to general concepts introduced already by Thomas Bayes. In contrast, Boltzmann's $\mu$-space distributions seem to represent objective states of many particles rather than incomplete information. In practice, however, these discrete distributions are always replaced by smooth ones, thus using some kind of coarse graining (neglect of information) in an essential way; any discrete distribution in a continuum would possess infinite negative entropy. So one may raise John Bell's fundamental question "Information by whom …?", while his other question "…about what?" seems to be answered by "about points in $\Gamma$-space". Unfortunately, this answer fails to explain the absence of factorials *N!* that would result from counting particle permutations – an important argument against the concept of particles (which are always distinguishable by their earlier positions).

The *physical* concept of entropy is usually defined by the finite "size" (volume in $\Gamma$-space) of an ensemble of microscopic states that may in some sense represent a given "macroscopic state". Its precise definition is in general not very important, since entropy is defined as a logarithmic measure, which means that an uncertainty by a factor of $X$ in the size of the ensemble would merely give rise to a relative correction of entropy of the order $\ln X/N$ (with $N$ of order *$10^{20}$*). Precision of ensemble measures does become important, though, for questions of principle, such as during measurements or in considerations regarding Maxwell's demon or Szilard's engine. The difference between a canonical and a micro-canonical ensemble, for example, is even physically meaningful, as only the former contains the entropy representing lacking information about energy fluctuations in open systems – and similarly for particle fluctuations in the grand canonical ensemble.

Microscopic determinism leads to the conservation of an appropriate size of an ensemble (its phase space volume). However, if physical entropy is understood as a function of *given* macroscopic properties, it is at most indirectly related to an ensemble that represents actual information held by an observer. In particular, physical entropy is defined as an extensive (additive) quantity, while conserved ensemble entropy is *not*, since it strongly depends on (dynamically arising) probability correlations between subsystems. This is the reason why Boltzmann's $\mu$-space entropy may increase even for deterministic collisions between particles; $\mu$-space information is thereby deterministically transformed into "irrelevant" information about (non-local) correlations.[9] In the case of irreversible phase transitions by formation of new order parameters, lacking information about thermal degrees of freedom (physical entropy) may even be transformed into lacking information about macroscopic variables (values of the order parameters). Hence, physical entropy can in principle be lowered if macroscopic properties are regarded as "given" as soon as they arise. Although this effect is usually negligible and essentially a matter of definition of physical entropy, the true surprise comes when one adds the (human?) observer to the chain of interacting systems in analogy to von Neumann's description of quantum observations. Deterministically, different macroscopic properties are then correlated with the arising different states of an observer. If the latter is assumed to "know" (be aware of) his own state, the ensemble describing his lack of knowledge would be reduced without violating microscopic determinism – in this way absolutely reducing the ensemble entropy from his point of view! Quantum-mechanically, this observation process of a microscopic quantity would require an indeterministic collapse of the wave function, but one may even discuss a *classical* version of *Wigner's friend*, who is known from quantum measurements, in this way. (In these dynamical descriptions of observations, the permanent and unavoidable formation of *uncontrollable* correlations such as by molecular collisions, which is the major source of irreversibility, is neglected for simplicity.)

This situation is obviously related to Maxwell's demon, who was suggested to use his presumed initial knowledge about molecular motions in order to reduce the thermal entropy by controlling a molecular trap door. Leo Szilard argued by means of a thought experiment that the demon's entropy must correspondingly increase if he is regarded as a physical object, too. Therefore, the experimenter, who would have to observe the gas molecules in order to act as a demon, must not be regarded as an extra-physical system. In Rolf Landauer's formulation: information is physical! Charles Bennett concluded[10] in accordance with traditional formulations of the second law that Maxwell's demon cannot work in a cyclic process that would be required for a perpetuum mobile of the second kind, because he would have to get rid of his information (and thereby raise the entropy of his environment) in order to close each cycle. However, lowering the ensemble entropy in an *individual* process as indicated above would nonetheless be possible in principle by interaction with the information-defining observer. On the other hand, any observer must already have got rid of quite a lot of entropy in order to come into being during a process of self-organization in an open system.

The (neg)entropy of information about macroscopic properties is usually negligible when quantitatively compared with thermodynamic entropy. For this reason, observers are in practice often regarded as extra-physical, or as possessing unlimited information capacity, and even defining their own arrow of time. This position becomes particularly problematic when applied to quantum states that were understood epistemically, that is, as representing a new kind of incomplete "quantum information" that is nowhere explicitly taken into account in a statistical definition of entropy. For consistency, any incomplete information must be represented by an ensemble of possible microscopic states. In a classical world, all measurement outcomes are in principle determined in advance by the global microscopic state (though in general not *known* in advance to an observer), and may then be "observed" (with or without being perturbed) rather than being *created* outside the laws of physics. While microscopic variables can usually be regarded as randomized before being measured, macroscopic ones are redundantly "documented" in their environment (for example by the light they have scattered, and that might later be received by observers). For this reason they appear to be "objectively given"; one cannot assume just *one* individual document (representing knowledge) to be different in order to change the past. This retarded multiple documentation[9] of macroscopic "facts" requires a strong time asymmetry of the physical world that establishes its "causal appearance" and an apparently fixed macroscopic past that is often simply disregarded in speculations about time travel and closed timelike curves in spacetime.

**3. The observer in the Everett interpretation**

Hugh Everett first recognized that in quantum theory we do not have to postulate a dynamical collapse of the wave function even if we require the wave function to describe reality. We may instead consistently assume that, after an observation, conscious observers exist in various independent "versions" $i$ that are the formal consequence of von Neumann's *unitary* description,

$$(1) \quad \left(\sum_i c_i \psi_i^S\right) \psi_0^A \psi_0^O \rightarrow \left(\sum_i c_i \psi_i^S \psi_i^A\right) \psi_0^O \rightarrow \sum_i c_i \psi_i^S \psi_i^A \psi_i^O =: \sum_i c_i \psi_i^{rel} \psi_i^O .$$

Its first step is sometimes called a "pre-measurement", while the suffixes *S, A,* and *O* indicate the system, apparatus and observer, respectively. Any information medium, such as transmitted light, is here for simplicity regarded as part of the apparatus. The states $\psi_i^{rel}$ on the right hand side can be identified with the potential final states of a stochastic collapse process that would according to the orthodox interpretation have to form before the outcome is observed. According to Everett, they define the "relative state" of the outside world with respect to each individual physical state of the subjective observer $\psi_i^O$ (therefore the title "Relative State Interpretation" of his original publication). The observer remains here somewhat vaguely defined, although, while remaining quite passive in (1), he evidently assumes a crucial role because of the kinematical quantum nonlocality (nonseparability).[11,12,13] In this description, the indeterminism that he observes is his own; it does not occur in the reality "out there". If unitary quantum dynamics applies universally, one cannot avoid such an entangled superposition that contains *different states* for all observers, who must then also be aware of different outcomes – just as different classical observers are aware of different things.[14] If some fundamental configuration space rather than common space is the stage for reality (that is, for the wave function), observers who are *parts* of the real world must be semi-autonomous objects in this "configuration" space. The question why there appears to be just one definite measurement outcome would therefore be similar to the question in classical cosmology why you, as an observer, are somebody rather than the whole world! Emphasis on this aspect of quantum observation has led to the name "many minds" or "multi-consciousness" interpretation, since the relative state with respect to the observer's mind describes the specific oberserver's "frog's perspective" of the quantum world – traditionally regarded as a dynamically resulting collapse component. Note that his passive role of the observer does not allow him to make any "decisions" or to "select" his future world branch by some free will.

Everett's conclusion was almost unanimously rejected by the physics community at its time for several reasons (if not just because of its unconventional nature). The major one was that the wave function had traditionally been regarded as meaningful only in the microscopic realm – at most until the Heisenberg cut is applied somewhere, whereupon it was assumed to "lose its meaning".[2] Those who did consider the general validity of the wave function as a possibility raised the objection that the expansion $\psi^{total} = \sum_i c_i \psi_i^{rel} \psi_i^O$ is defined with respect to any basis of states $\psi_i^O$ chosen for the observer system (including states which would represent superpositions of *different* states of awareness). In von Neumann's first step of Equ. (1), the *i*-basis is usually defined by means of a phenomenological "observable" that is used to characterize a specific measurement – according to Bohr regardless of whether an observer ever enters the scene in order to read off the result. This basis was later called the "pointer basis".[15] If the observer system $O$ at the end of the observational chain of interactions was in principle precisely defined, one could use the essentially unique Schmidt canonical representation, which is defined as a single sum for both subsystems, to define a subjective observer basis. However, this representation would fluctuate in time and strongly depend on the precise boundary between this observer system $O$ and the rest of the quantum world.[16] Although it may be fundamentally important, it is not particularly appropriate to describe objective measurements by means of an apparatus or macroscopic memory device.

The problem of how to justify an objective pointer basis that would specify an apparent ensemble of outcomes in a measurement (but *not* the pseudo-problem of obtaining one definite outcome) was resolved by the theory of environmental decoherence.[17] This very gen-eral phenomenon had not been correctly recognized for a long time, since it is also based on the assumption, used in (1), that unitarity is valid beyond microscopic systems. It leads to the further consequence that the relative states in (1) have to include an uncontrollable and normally inaccessible environment of the macroscopic apparatus, that is,

$$\psi_i^{rel} = \psi_i^S \psi_i^A \psi_i^{env} \quad . \tag{2}$$

Because of the unavoidably arising essential *i*-dependence of the environmental states $\psi_i^{env}$, a superposition of macroscopically different states $\psi_i^A$, formed in a unitary measurement, is immediately and irreversibly "dislocalized" over many degrees of freedom, and thus inaccessible to local observers with their local interactions. In this way, the "normal" and usually unavoidable environment of a macroscopic system induces a preferred basis for the pointer variable or any other quasi-classical property that is objectively characterized by its robustness against further decoherence. The experimental verification of decoherence was in fact historically the first confirmation that entangled wave functions are relevant not only for microscopic systems, while this entanglement must then be far more complex than envisioned by von Neumann and Everett with their simplified models. The *i*-dependent effect in the environment does here *not* necessarily have to represent any usable information or documentation, since even an (initially separate) thermal environment is sufficient to cause decoherence. Therefore, only a minority of quasi-classical variables may be assumed to be "physically given" (macroscopic) in the sense of Sect. 2. Although it remains conceivable that the unbounded dislocalization of superpositions (decoherence) will at some point lead to an instability caused by some very small and as yet unknown deviation from global unitarity that would define a collapse, this would then not amount to any observable difference: all *observed* quantum jumps would still have to be described by a decoherence process.

Because of the locality (in space) of all interactions, the preferred basis is usually the position basis of a pointer or another macroscopic variable – thus giving rise to the appearance of particles, for example. Although a reversal of this dislocalization of superpositions would in principle be compatible with the Schrödinger equation, it is again excluded by the arrow of time characterizing our world (regarded as a fact rather than a law – namely the initial absence of any correlations or entanglement which might have local effects at any reasonable later times). Nonlocal superpositions that are "caused" according to (1) by taking into account the environment can thus not be relocalized any more, although they never disappear from the universe unless the Schrödinger dynamics was modified. Therefore, any description of reality in terms of a unitarily evolving wave function requires an Everett interpretation – the reason why I proposed and mentioned it, originally without knowing of Everett, when I prepared my first paper on decoherence (that would appear in 1970).[17]

Realistically, the macroscopic "apparatus" that leads to an observation in (1) would not only have to include the thereby required information medium or registration device, but also the human sensory organs and much of his neuronal system. Both are macroscopic in the sense of being decohered *and* documented in their environments,[18] and both are – partly, at least – external to any reasonable subjective observer system (the physical carrier of consciousness in the brain). Although the neuronal apparatus is indeed a particularly fine-grained (complex) system whose variables have to be assumed to be always "given", its further decoherence after a measurement does not affect the latter's objective result in any way. The precise localization of consciousness in the brain remains an open problem – just as it did in all classical descriptions, although one may expect that, in contrast to controllable information, it has ultimately to be described in quantum mechanical (or entirely novel) terms.

After decoherence by the environment, the macroscopic system may for all practical purposes be characterized by its reduced density matrix, such as $\rho_{red}^A = \sum_i |c_i|^2 \psi_i^A \psi_i^{A*}$. Although this density matrix may be identical to that which would represent an ensemble of states postulated by a collapse, it does by its very definition as a partial trace *not* represent an ensemble. If it did, a subsequent observation of the property *i* (an "increase of information") would simply cause the observer state thereafter to depend on the already "given" value of *i*. In unitary description, however, the observation of an individual outcome can only be a consequence of Everett's splitting observer,[11,14,16,17] since the global superposition (1) that includes the environment, and that formally gave rise to the density matrix for subsystems, now consists of various dynamically autonomous world components which describe different macroscopic properties and observers. Therefore, the different observer states $\psi_i^O$, whatever their precise definition, must approximately be factors of the decohered branch states, and can thus be no more aware of the existence of the "other worlds" that are described by the other component states $\psi_{i \neq i'}^{rest}$.

This consequence is sufficient for the theory to consistently describe all our observations in an apparently classical quantum universe. Note that the molecules forming a gas, for example, are also decohered into narrow wave packets by their mutual collisions, and thus effectively define a quasi-classical $\mu$-space distribution, but this does *not* justify quasi-deterministic trajectories for them. Their collisions would appear stochastic in such a quasi-classical description (thus precisely justifying Boltzmann's *Stosszahlansatz*). For this reason, their positions, although "preferred" by decoherence, cannot be assumed to be macroscopic and "given" in each Everett branch;[9] they do not represent accessible "information".

Since the dynamical autonomy of Everett branches has thus been clearly established by decoherence, the localization of the subjective observer system in states existing within these branches appears indeed so plausible or "normal" that, for example, some Oxford quantum philosophers[19] regard the quantum measurement problem as solved by the combination of decoherence and Everett without explicitly mentioning the observer – again in order to save the concept of an observer-independent reality. Nonetheless, the physical specification of the observer with his "many minds" is an important element of a quantum theory of observation. It can only be justified by the locality of the subjective observer (that is in turn a consequence of the *dynamical* locality of this kinematically *nonlocal* quantum world[11]).

As an example, consider two spatially separated microscopic systems entangled with one another as in a Bell state. If one of them is locally measured, *both* get immediately entangled with the apparatus and its local environment – but nothing else as yet. An observer at the location of the other microscopic system, say, will participate in the entanglement only after having received a signal about the outcome. Only thereafter will he be in different states in the different "worlds" that became dynamically separated from one another by the irreversible decoherence of the pointer position, but which together still form but *one* quantum world. If he decides to measure and observe also the microscopic system at his own location (before or after receiving the first signal – thus including delayed choice experiments), his state splits further in order to separately register and become aware of the quasi-classical outcomes of *both* measurements. When repeating the total measurement many times, he would in "most" of his resulting versions in very good approximation confirm the frequencies predicted by Born's rule (and thus their statistical correlations that violate Bell's inequality) – provided the branches containing his various versions possess statistical weights according to their squared norms. Any other conceivable statistical weights that would, in contrast, *not* be conserved under the Schrödinger equation (such as an ill-defined *number* of branches) would lead to probabilities that might later change under further branchings that happen asymmetrically in *some* branches only (thus asymmetrically increasing their number, but not their total norm). Everett considered this consequence as a proof of Born's probability measure[20] – although it is no more than a plausibility argument for this statistical postulate that regards the passive observer rather than "reality".

All those "weird consequences" of quantum mechanics that have been "discovered" and much discussed in the media during recent decades can similarly and consistently be described in wave mechanical terms in a high-dimensional configuration space, since this is indeed how most of them were predicted. Their apparent weirdness is merely a consequence of the traditional tendency to describe the world in classical (local) terms. The reader may himself analyze the so-called quantum teleportation protocol as a second example in purely quantum mechanical terms in order to confirm that everything that appears to be teleported (or its local causal predecessor) must have been prepared in advance by subluminal means at its target position in at least one of the components of the required entangled wave function.[21] Teleportation and other "esoteric" phenomena would instead be required if local properties came into existence only in corresponding measurements (as assumed in the Copenhagen interpretation). It becomes again evident in this way that entanglement cannot merely represent statistical correlations, even though one may *pretend* that an initial superposition is transformed into one member of an ensemble that would then represent lacking information as soon as decoherence has become irreversible in a chain of interactions that might lead to an observation. (The cat has to be assumed to have died – if it ever did so in the corresponding "world" – long before the box was opened.) As this *pretended* collapse is not a *physical* process, it may even be defined to "occur" superluminally, although such unphysical dynamics can obviously not be used to send information. This restriction of the apparent quantum reality to one (not yet known) effective branch wave function by a collapse at the time when decoherence has become irreversible is certainly convenient, and thus pragmatically justified, but physics students should be taught to understand its correct meaning in a consistent theory.

Any hypothetical proposal for a *genuine* (physical) collapse would have to be specified in order to be meaningful, whereby it has to avoid inconsistencies with the principles of relativity. A (conceivable) empirical verification of such a violation of the Schrödinger equation might then falsify Everett's interpretation, while an *unspecified* collapse proposal can hardly ever be falsified. Most collapse models in the literature still contain free parameters that would also leave them non-falsifiable as long as these parameters do not have to obey certain bounds in order to fulfill their purpose of predicting definite measurement outcomes. Therefore, the dispute about a collapse of the wave function in contrast to Many Worlds should not be a matter of "religious" belief or prejudice. There simply exist two classes of possibilities whose consequences should be further analyzed and tested, while the original Copenhagen interpretation with its fundamental classical concepts seems to be deprived of its major motivation by the success of the decoherence program. The latter argument also applies to Bohm trajectories. If we can explain the *appearance* of quasi-classical properties, we do not have to assume any further (hidden) reality for them, such as local "beables". It should also be obvious that the wave function can carry information only if it represents a physical (real) object.

The consistent description of quantum phenomena according to Everett's interpretation thus means that the observed quantum indeterminism does not represent a stochastic dynamical process in nature, since the global wave function is assumed to evolve deterministically. Rather, it reflects the multiple future of an observer in this deterministic quantum world – comparable to a process of cell division in a deterministic classical world. Without a subjective observer, there would be no justification for a frog's perspective that is related to branches of the wave function. However, this indeterminism of the observer's history is objectivized with respect to those versions of other observers (including "Wigner's friends") that are correlated by their entanglement. Since they are defined to "live" in the same world branch, they agree about the outcomes of measurements, while all their other versions do not have to disappear from reality; it is sufficient that they cannot communicate any more with those in our branch. This entanglement between different observers is the same as that between an observer and his apparatus according to Eq. (1).

So we may have to conclude that the Everett interpretation, which is a direct consequence of the Schrödinger equation (and thus falsifiable in principle by the conceivable discovery of a limitation of its validity), is indistinguishable *in practice* from the pragmatic although never precisely defined Copenhagen interpretation. In contrast to the latter it is conceptually consistent, and thus compatible with the concept of a well-defined (though kinematically nonlocal) micro-physical reality. In particular, it avoids all irrational concepts such as spooky action, complementarity, and any "uncertainty" of its basic kinematical terms – and so does not offer any justification for speculations about supernatural or extra-physical phenomena. However, the possibility that the historical selection of *our* Everett branch may require some *improbable* evolutionary events in accordance with the weak anthropic principle cannot be excluded.